\documentclass[%
reprint,
amsmath,amssymb,
aps,
]{revtex4-2}

\usepackage{graphicx}%
\usepackage{dcolumn}%
\usepackage{bm}%
\usepackage{hyperref}%

\usepackage{graphicx}	%
\usepackage{amsmath}	%
\usepackage{bm}
\usepackage{glossaries} %
\usepackage{color}
\usepackage{xspace}
\usepackage{booktabs}

\newcommand{\days}{\ensuremath{\mathrm{days}}\xspace}

\newcommand{\wfi}{\mbox{WFI J2033 -- 4723}\xspace}
\newcommand{\bone}{\mbox{B 1608 + 656}\xspace}
\newcommand{\he}{\mbox{HE 0435 -- 1223}\xspace}

\setacronymstyle{long-short} %
\glsdisablehyper %
\newacronym{agn}{AGN}{active galactic nuclei}
\newacronym[plural={GPs}, longplural={Gaussian Processes}]{gp}{GP}{Gaussian Process}
\newacronym{gpr}{GPR}{Gaussian Process regression}
\newacronym{carma}{CARMA}{continuous auto-regressive moving average}
\newacronym{ns}{NS}{Nested Sampling}
\newacronym{kl}{KL}{Kullback-Leibler}
\newacronym{kde}{KDE}{Kernel Density Estimate}
\newacronym{sm}{SM}{Spectral Mixture Kernel}

\newacronym{e}{E}{Exponential}
\newacronym{m32}{M32}{Mat\'ern-3/2}
\newacronym{m52}{M52}{Mat\'ern-5/2}
\newacronym{m72}{M72}{Mat\'ern-7/2}
\newacronym{se}{SE}{Squared Exponential}
\newacronym{rq}{RQ}{Rational Quadratic}

\newcommand{\timedelayval}[2]{$#1 \pm #2$}

\newcommand{\missingdata}{\textit{---}} %

\begin{document}
	
\preprint{APS/123-QED}

\title{Time delays and stationarity in quasar light curves}

\author{Namu Kroupa$^{1,2,3}$}
\email{nk544@cam.ac.uk}
\author{David Yallup$^{2,4}$}
\author{Will Handley$^{2,4}$}
\affiliation{$^1$Astrophysics Group, Cavendish Laboratory, J.J. Thomson Avenue, Cambridge, CB3 0HE, UK}
\affiliation{$^2$Kavli Institute for Cosmology, Madingley Road, Cambridge, CB3 0HA, UK} 
\affiliation{$^3$Engineering Laboratory, University of Cambridge, Cambridge CB2 1PZ, UK}
\affiliation{$^4$Institute of Astronomy, University of Cambridge, Madingley Road CB3 0HA, UK}

\date{\today}

\begin{abstract}

We present a fully Bayesian framework for time delay inference and stationarity tests in quasar light curves using marginalised Gaussian processes. The model separates a deterministic, non-stationary drift (piecewise linear mean) from stationary stochastic variability (Mat\'ern and Spectral Mixture kernels), and jointly models multiple images with per-image microlensing. Bayesian evidence and parameter posteriors are obtained via nested sampling and marginalised over model choices.
Applied to the quasars \wfi, \bone, and \he, we find strong evidence for non-stationarity in \bone and \he, while \wfi is consistent with stationarity. The stochastic component favours an Markovian exponential kernel for \bone and a non-Markovian Mat\'ern-$\frac32$ kernel for \wfi and \he. Multi-length-scale Spectral Mixture kernels are disfavoured. 
Time delays are shown to be robust to model assumptions and consistent with prior work within the error. 
We further identify and mitigate a likelihood pathology which biases toward large delays, providing a practical nested sampling convergence protocol.
\end{abstract}

\maketitle

\section{Introduction}

The study of quasars, a particularly bright class of \gls{agn}, serves as an important probe for understanding galaxy evolution and cosmology~\citep{york2000sloan, shen2011catalog}. A defining characteristic is their pronounced variability across the electromagnetic spectrum, observed over a vast range of timescales, from minutes to decades~\citep{ulrich1997variability, vaughan2013random, ozdonmez2024multi}. This variability is intrinsically linked to the physical processes occurring in the vicinity of the supermassive black hole, including instabilities in the accretion disk, thermal fluctuations, and the reprocessing of radiation, providing information about the accretion disc, the supermassive black hole properties, and the circumnuclear environment~\citep{kelly, padovani2017active, peterson2001variability}. 

The analysis of quasar light curves, therefore, provides a method for investigating these unresolved regions and testing accretion theories, but presents several difficulties. Ground-based surveys often produce irregularly sampled data with seasonal gaps, complicating standard time series analysis~\citep{covino2022detecting, chen2024searching}. Photometric measurements also include noise, which can be confused with low-amplitude intrinsic variability. Quasar variability is largely stochastic, often characterized as red noise, with greater power on longer timescales~\citep{kelly, macleod2010modeling}. This stochasticity makes it challenging to identify deterministic signals like periodicities. Moreover, quasar light curves exhibit a wide spectrum of complex behaviours, including long-term trends and distinct patterns that are not fully captured by simple stationary stochastic models~\citep{luo2020characterization, stone2022optical}. These observations suggest the influence of non-stationarity in the underlying physical processes or the combined effect of multiple interacting mechanisms.

The challenge of accurately estimating time delays from complex, noisy, and irregularly sampled light curves is central not only to understanding intrinsic \gls{agn} physics but also to their cosmological applications. Time delay cosmography, which utilises the time delays between multiple images of a gravitationally lensed variable quasar to measure absolute distances and infer the Hubble parameter $H_0$~\citep{refsdal1964possibility, treu}, critically depends on robust delay measurements. Within the context of the Hubble tension~\citep{shah, abdalla2022cosmology}, time-delay cosmography provides precise and accurate $H_0$ values~\citep{suyu2013two, wong2020h0licow, millon2020tdcosmo, birrer2020tdcosmo}. However, observational challenges, including distortions from microlensing, persist~\citep{hojjati2013robust, hojjati2014next, tewes2013cosmograil}. A statistically analogous problem to gravitational lens time delay estimation is faced by reverberation mapping, used to probe the broad-line region and accretion disc sizes~\citep{blandford1982reverberation, peterson1993reverberation, zu2011alternative, kelly2014flexible, mcdougall2025litmus}.

Traditional Fourier-based methods for time series analysis, such as those based on the power spectral density or the cross-spectrum, struggle with the irregular sampling typical of astronomical data~\citep{vaughan1997x, uttley2002measuring, epitropakis2016statistical, vaughan2013random}. While localised variability can be identified using methods like Bayesian Blocks~\citep{scargle2013studies} and machine learning offers new approaches~\citep{tachibana2020deep, fagin2024latent}, stochastic process models remain a cornerstone. The damped random walk model gained popularity for its simplicity in describing quasar optical variability~\citep{kelly, kozlowski2009quantifying, macleod2010modeling}. The empirical linear \emph{rms-flux} relation~\citep{uttley2001flux} and the associated lognormal flux distribution of quasars found theoretical backing in the damped random walk framework, where the flux is given by $x(t)=\exp[\ell(t)]$ with $\ell(t)$ being a linear Gaussian stationary stochastic process~\cite{uttley2005non}. However, deviations from damped random walk predictions, particularly steeper power spectral density slopes at short timescales~\citep{mushotzky2011kepler, zu} and non-stationarity in highly variable \gls{agn} like \mbox{IRAS 13224 -- 3809}~\cite{alston2019remarkable}, evidenced by non-lognormal flux distributions and non-linear rms-flux relations~\citep{alston2019non, bhattacharyya2020blazar}, necessitate more flexible models. Knowing whether a stochastic process is stationary is crucial for linking variability to physical models and understanding connections between emission components~\citep{alston2019remarkable}. 

\glspl{gp} provide a non-parametric Bayesian framework for modelling such time series, naturally handling irregular sampling and offering principled uncertainty quantification~\citep{rasmussen, roberts, hojjati2013robust, kelly2014flexible}. The choice of the \gls{gp} kernel and its hyperparameters is critical. Traditional maximum likelihood estimation can be problematic for complex likelihood surfaces or influential priors. Furthermore, pre-selecting a single kernel introduces strong biases, especially for extrapolation tasks like time delay estimation, where stationary kernels with simple mean functions revert to the prior mean far from data~\citep{kroupa2024kernel, kroupa2026global}. 
The \gls{gp} approach also avoids the insensitivity of binned data methods to non-stationary variations on timescales shorter than the bin width.

A standard frequentist hypothesis test for stationarity, the Augmented Dickey-Fuller test~\citep{said1984testing, bhattacharyya2020blazar}, has its own limitations. Firstly, it allows only pairwise comparison between a specific non-stationary and stationary model. Reformulated in terms of \glspl{gp}, the null hypothesis may be rewritten as possessing a polynomial mean function with a non-stationary kernel, while the alternative hypothesis possesses a mean function which is a polynomial of one order lower and a stationary kernel. When extrapolating a time series, which is strictly necessary for time delay inference~\cite{kroupa2026global}, polynomial mean functions may lead to unphysical diverging light curve magnitudes unless higher order coefficients are heavily regularised. Secondly, the precise form of the \gls{gp} kernel in each hypothesis is set by the number of time lags included in the test, which can itself be determined by an Akaike information criterion or similar. However, as observed previously~\citep{warnes1987problems, lalchand2020approximate, kroupa2024kernel}, Gaussian Process posterior distribution may deviate strongly from the assumption of Gaussianity underlying such information criteria and hence bias the inference.

As a Bayesian method, \glspl{gp} are amenable to extension in a fully Bayesian framework 
by marginalising over kernel and mean function choices to improve robustness~\citep{kroupa2024kernel}. 
This paper focuses on flexible, data-driven mean functions parameterised as piecewise linear splines with flexible knots~\citep{vazquez2012reconstruction, hee, ormondroyd2025nonparametric} combined with stationary kernels from the Mat\'ern and \gls{sm} families.
Especially, we do not rely on time series descriptors, which are ultimately proxies for underlying stochastic processes~\citep{scargle2020studies}. Finite data, especially when the observation length is insufficient relative to intrinsic timescales, limits inference~\citep{zhang2004inconsistent, velandia2017maximum, kroupa2024kernel}. We therefore use Bayesian inference to make probabilistic statements about model comparison.
This approach of marginalising over the mean function differs from and aims to be more robust than \gls{gp} applications with fixed, simple mean functions \citep{hojjati2013robust}. The potential to link \gls{gp} kernel parameters to underlying quasar properties \citep{covino2022detecting, chen2024searching} further motivates the development of robust kernel selection and mean function modelling. Time delays between different wavebands, another key observable related to disc reprocessing, also stand to benefit from such improved modelling.

In this paper, we analyse the light curves of the quadruply lensed quasars \wfi~\citep{vuissoz2008cosmograil}, \bone~\citep{fassnacht2002determination} and \he~\citep{bonvin}. 
These objects represent well-studied benchmark systems for time delay cosmography and microlensing studies due to their unique physical properties and the availability of long term, high-quality monitoring data, which make them ideal for Gaussian process modelling. 

\wfi~\cite{morgan2004wfi, vuissoz2008cosmograil} is distinguished by a close pair of lensed images and exhibits significant microlensing, providing a valuable dataset for constraining quasar accretion disc sizes~\cite{blackburne2011sizes}. Long-term monitoring, notably by the COSMOGRAIL program~\cite{millon2020cosmograil}, has yielded precise time delays for this system. In contrast, the radio-selected lens \bone~\cite{myers19951608+} features a complex environment of two interacting lens galaxies and is largely free of microlensing in radio observations. This has made it a crucial target for clean measurements of intrinsic quasar variability and a foundational system for determining $H_0$ through VLA monitoring~\cite{fassnacht2002determination, koopmans2003hubble, kochanek2004tests, suyu2010dissecting}. Finally, \he~\cite{wisotzki2002he}
is a bright quad with a well-resolved Einstein ring and prominent microlensing~\cite{blackburne2011sizes}. It has become a cornerstone for $H_0$ measurements by the H0LiCOW~\cite{suyu2017h0licow, sluse2017h0licow, rusu2017h0licow, wong2017h0licow, bonvin2017h0licow} and TDCOSMO~\cite{millon2020tdcosmo, birrer2020tdcosmo} collaborations, owing to extensive and precise time delay data from COSMOGRAIL~\cite{courbin2011cosmograil}. 

The multi-decade and densely sampled light curves for these three objects provide multiple time-shifted views of the same intrinsic signal. This wealth of data allows for the robust separation of intrinsic quasar variability from extrinsic microlensing effects, setting them apart from typical quasar observations and enabling precise astrophysical measurements.

The structure of this paper is as follows. Section~\ref{sec:gps} introduces the flux variability model. 
Section~\ref{sec:real-data-method} describes the adopted protocol for sampling the \gls{gp} model as applied to the light curves of \wfi~\citep{vuissoz2008cosmograil}, \bone~\citep{fassnacht2002determination} and \he~\citep{bonvin}. The inference results are explained in Section~\ref{sec:results}. Finally, the conclusions are provided in Section~\ref{sec:conclusions}.

\section{Background}~\label{sec:gps}

In this section, we first explain the modelling choices for the quasar light curves and consequently describe the concrete flux variability model implemented. Finally, we describe how separate models are combined with marginalised \glspl{gp}.

\subsection{Time delayed Gaussian processes}~\label{sec:time-delayed-gaussian-processes}

We model quasar light curves with time delayed \glspl{gp}.
In the standard \gls{gp} modelling paradigm~\cite{rasmussen}, these are defined with a zero-mean function and a stationary kernel and tend to revert to their prior mean when predicting far from the observed data. 
This characteristic hinders their ability to extrapolate, which is crucial when dealing with significant time delays where the function and its shifted version become effectively uncorrelated~\cite{kroupa2026global}. This limitation can lead to pathological structures in the posterior distribution of the time delay, creating practical issues for inference, as was recently established as a primary challenge for time delay inference.

To improve extrapolation, we are therefore required to choose non-stationary \glspl{gp}. For this, we may either choose non-stationary kernels, which do not decay over a certain length scale, or a non-constant mean function. Due to the success of the damped random walk model and related stochastic processes~\citep{kelly} and motivated by previous modelling of quasar light curves with splines~\citep{millon2020pycs3} or B\'ezier curves~\citep{luo2020characterization}, we choose to retain a stationary kernel and proceed with a non-constant mean function. 
One possibility to proceed is to use the reformulation of a spline
as the posterior mean of a \gls{gp}~\citep{wahba1978improper, schoenberg1988spline, mackay-2}. However, this 
formulation 
restricts the knots of the spline to lie on the data points. A less restrictive method of spline regression are \emph{flexknots}~\citep{vazquez2012reconstruction, hee, handley2019bayesian, vazquez2012model, aslanyan2014knotted, finelli2018exploring, ade2014planck, ade2016planck, millea2018cosmic, heimersheim2022takes, olamaie2018free, heimersheim2024flexknot, shen2024flexknot, ormondroyd2025nonparametric, ormondroyd2025comparison}, which promotes the knots to parameters which can be sampled.

A final modelling choice lies in the order of the splines. While in principle this is a choice which can be marginalised over, we choose linear splines as this confines the effect of each linear segment of the spline to its own interval and does not induce correlations between data points far apart, compared to cubic splines for instance. This is desirable from an interpretability standpoint. 

In summary, we choose stationary kernels and a mean function parameterised by a linear spline with flexible knots. In the next section, we concretely describe the adopted \gls{gp} model for the flux variability.

\subsection{Flux variability model}

Following~\citet{hojjati2013robust}, we assume that all observed light curves are instances of a latent \gls{gp} but with shifts in magnitude and time, additional noise and a distortion from microlensing, also modelled as a \gls{gp}. 
Departing from the standard \gls{gp} modelling paradigm, we use parametric models for long range variations and non-parametric models to model variations over short scales. This separation of scales is obtained in a purely data-driven manner.

In the following, we describe the \gls{gp} model for the case of $n_\mathrm{lc}=3$ light curves, A, B and C, with magnitudes $\mathbf{y}_\mathrm{A}$, $\mathbf{y}_\mathrm{B}$ and $\mathbf{y}_\mathrm{C}$, respectively, as the generalization to more light curves is straightforward.
From here on, we use the notation $\mathbf{f}(\mathbf{t})$ to denote the values of the function $f$ at the observation times $\mathbf{t}=[t_1,\dots,t_{n_\mathrm{data}}]^\top$ stacked into a column vector. Likewise, we denote a matrix evaluated at all pairs of observation times by ${\mathbf{K}(\mathbf{t},\mathbf{t})_{ij}=k(t_i,t_j)}$.

\subsubsection{Mean function}\label{sec:flux-model-mean-function}

We set the mean function of light curve A to ${m_\mathrm{A}(t)=\bar{\mathbf{y}}_\mathrm{A}+m_\mathrm{fk}(t)}$
by calculating the average $\bar{\mathbf{y}}_\mathrm{A}$ of $\mathbf{y}_\mathrm{A}$ and adding the flexknot $m_{\mathrm{fk}}$, which we define further below. At the observation times $\mathbf{t}$, the values of the mean function for light curve A are thus
\begin{equation}
\mathbf{m}_\mathrm{A}=\bar{\mathbf{y}}_\mathrm{A}\mathbf{1}+\mathbf{m}_{\mathrm{fk}}(\mathbf{t}),
\end{equation} 
where $\mathbf{1}=[1,1,\dots,1]^\top$.
The mean functions of the other light curves differ from $\mathbf{m}_\mathrm{A}$ by a relative offset in magnitude and a shift in time,
\begin{align}
\mathbf{m}_\mathrm{B}&=\mathbf{m}_\mathrm{A}(\mathbf{t}+\Delta t_{\mathrm{AB}}\mathbf{1})+\Delta m_\mathrm{AB}\mathbf{1},\\
\mathbf{m}_\mathrm{C}&=\mathbf{m}_\mathrm{A}(\mathbf{t}+\Delta t_{\mathrm{AC}}\mathbf{1})+\Delta m_\mathrm{AC}\mathbf{1}, 
\end{align}
where
$\Delta m_\mathrm{AB}$ and $\Delta m_\mathrm{AC}$ are the constant offsets of light curve B and C with respect to A. 
The light curves are therefore shifted instances of the latent \gls{gp} with mean function $\mathbf{m}_\mathrm{fk}$. 
We are thus using light curve A as a reference light curve and our mean function is equivalent to subtracting the average magnitude of light curve A from all light curves and modelling a latent \gls{gp}.
Therefore, we set uniform priors on $\Delta m_\mathrm{AB}$ and $\Delta m_\mathrm{AC}$ with range given by the maximum extent of the light curve shifted by light curve A, ${[\min_i(\mathbf{y}_j-\bar{\mathbf{y}}_\mathrm{A})_i,\max_i(\mathbf{y}_j-\bar{\mathbf{y}}_\mathrm{A})_i]}$ for light curve $j\in\{\mathrm{B},\mathrm{C}\}$.

\subsubsection{Flexknots}\label{section:flexknots}

We follow the implementation in~\citet{ormondroyd2025nonparametric}. 
A flexknot is a linear spline whose knots $(t_n^{(\mathrm{fk})},y_n^{(\mathrm{fk})})_{n=1}^{n_\mathrm{fk}}$ are parameters. The number of knots~$n_\mathrm{fk}$ is treated as a hyperparameter (Section~\ref{sec:marginalised-gps}). The $t$-values of the first ($t_1^{(\mathrm{fk})}$) and last ($t_{n_{\mathrm{fk}}}^{(\mathrm{fk})}$) knot are fixed to the $t$-values of the first ($t_1$) and last ($t_{n_\mathrm{data}}$) data points, respectively. A sorted-uniform prior~\cite{buscicchio2019label, handley} is set on the remaining $n_\mathrm{fk}-1$ $t$-values $(t_n^{(\mathrm{fk})})_{n=2}^{n_\mathrm{fk}-1}$. The sorted-uniform prior removes redundant regions of the parameter space given by relabelling of the knots. A uniform prior is set on the $n_\mathrm{fk}$ $y$-values. We take the convention that the spline continues to extrapolate linearly to the left and right. So far, the flexknot has $2n_\mathrm{fk}-2$ parameters with $n_\mathrm{fk}\ge 2$.
The value $n_\mathrm{fk}=0$ is now defined as the zero function and $n_\mathrm{fk}=1$ is set to the constant function with a uniform prior set on the constant $c_\mathrm{fk}$. In summary, the flexknot now has $2n_\mathrm{fk}-1$ parameters with $n_\mathrm{fk}\ge 0$ and is defined by
\begin{equation}
	m_\mathrm{fk}(t)=
	\begin{cases}
		0&\text{if }n_\mathrm{fk}=0,\\
		c_\mathrm{fk}&\text{if }n_\mathrm{fk}=1,\\
		\text{spline}_{(t_n^{(\mathrm{fk})},y_n^{(\mathrm{fk})})_{n=1}^{n_\mathrm{fk}}}(t)&\text{if }n_\mathrm{fk}\ge 2.
	\end{cases}
\end{equation}
Increasing $n_\mathrm{fk}$ therefore increases the model complexity.

\subsubsection{Covariance matrix}

The covariance matrix 
takes a block form. The diagonal blocks account for the correlations within each light curve,
\begin{align}
\mathbf{K}_\mathrm{A}&=\mathbf{K}(\mathbf{t},\mathbf{t}),\\ \mathbf{K}_\mathrm{B}&=\mathbf{K}(\mathbf{t}+\Delta t_\mathrm{AB}\mathbf{1},\mathbf{t}+\Delta t_\mathrm{AB}\mathbf{1}),\\ 
\mathbf{K}_\mathrm{C}&=\mathbf{K}(\mathbf{t}+\Delta t_\mathrm{AC}\mathbf{1},\mathbf{t}+\Delta t_\mathrm{AC}\mathbf{1}). 
\end{align}
The off-diagonal blocks represent the correlation between different light curves,
\begin{align}
\mathbf{K}_\mathrm{AB}&=\mathbf{K}(\mathbf{t},\mathbf{t}+\Delta t_\mathrm{AB}\mathbf{1}),\\ \mathbf{K}_\mathrm{AC}&=\mathbf{K}(\mathbf{t},\mathbf{t}+\Delta t_\mathrm{AC}\mathbf{1}),\\
\mathbf{K}_\mathrm{BC}&=\mathbf{K}(\mathbf{t}+\Delta t_\mathrm{AB}\mathbf{1},\mathbf{t}+\Delta t_\mathrm{AC}\mathbf{1}).
\end{align}
We choose the kernel function from the family of Mat\'ern kernels (\gls{e}, \gls{se}, \gls{m32}, \gls{m52}, \gls{m72}), the \gls{rq} kernel and Spectral Mixture (SM) kernels. These choices are defined and discussed in Appendix~\ref{appendix:kernel-choices}. The choice itself is treated as a hyperparameter (Section~\ref{sec:marginalised-gps}). 
Distortion from microlensing on each light curve is modelled with a \gls{gp} with a \gls{se} kernel. 
Hence, the microlensing term $\mathbf{K}_{\mu}$ is added to each block diagonal of the covariance matrix.

Finally, we include the error bars on the magnitude measurements, $\bm{\sigma}_\mathrm{A}$, $\bm{\sigma}_\mathrm{B}$ and $\bm{\sigma}_\mathrm{C}$, as well as one adjustable white noise parameter $\sigma$ so that the noise terms are
\begin{align}
\bm{\Sigma}_\mathrm{A}&=\mathrm{diag}(\bm{\sigma}_\mathrm{A})^2+\sigma^2\mathbf{I},\\ \bm{\Sigma}_\mathrm{B}&=\mathrm{diag}(\bm{\sigma}_\mathrm{B})^2+\sigma^2\mathbf{I},\\ \bm{\Sigma}_\mathrm{C}&=\mathrm{diag}(\bm{\sigma}_\mathrm{C})^2+\sigma^2\mathbf{I},
\end{align}
where 
$\mathrm{diag}(\bm{\sigma}_\mathrm{A})$ places the entries of $\bm{\sigma}_\mathrm{A}$ on the diagonal of a matrix and $\mathbf{I}$ is the identity matrix.

Uniform priors are set on all covariance matrix parameters.
Priors on amplitude parameters (kernel amplitude~$A$, \gls{sm} kernel weights~$w_q$, microlensing kernel amplitude~$A_\mu$) range from zero to the square root of half of the light curve magnitude range~\cite{kroupa2024kernel}. Length scale priors (kernel length scale $\ell$, \gls{sm} kernel scale $\ell_q$, microlensing kernel length scale $\ell_\mu$) are set from zero to the difference between the largest and smallest observation times of the data set, $\Delta t_\mathrm{range}=t_{n_\mathrm{data}}-t_1$.
For the \gls{rq} parameter $\alpha$, it was generically observed~\cite{kroupa2024kernel} that most of the posterior mass, covered by more than 5 standard deviations around the mean, is roughly within the interval $[0,10]$ so that we use this range for the prior bounds here.
The prior on each \gls{sm} kernel frequency $f_q$ is set to range from zero to the Nyquist frequency $f_\mathrm{Nyquist}=\frac{n_\mathrm{data}}{2\Delta t_\mathrm{range}}$.
The white noise term $\sigma$ is set to range from zero to the maximum error bar among all light curves, $\max_{i,j}(\bm{\sigma}_i)_j$.

In practice, we observe that the bulk of the posterior mass is well contained within these prior ranges, except for the length scale of the \gls{e} kernel, which is attributed to the data and not an intrinsic feature of the likelihood~\cite{kroupa2026global}.

\subsubsection{Full Gaussian process model}

The magnitudes $\mathbf{y}_\mathrm{A}$, $\mathbf{y}_\mathrm{B}$ and $\mathbf{y}_\mathrm{C}$
are stacked into a single column vector.
Combining the individual light curves in such a way is known as a multi-output \gls{gp}:
\begin{equation}
	\begin{bmatrix}
		\mathbf{y}_\mathrm{A}\\
		\mathbf{y}_\mathrm{B}\\
		\mathbf{y}_\mathrm{C}
	\end{bmatrix}
	\sim
	\mathcal{N}
	\left(
	\begin{bmatrix}
		\mathbf{m}_\mathrm{A}\\
		\mathbf{m}_\mathrm{B}\\
		\mathbf{m}_\mathrm{C}
	\end{bmatrix},\mathbf{K}
	\right)
\end{equation}
with covariance matrix
\begin{equation}
	\mathbf{K}=\begin{bmatrix}
		\mathbf{K}_\mathrm{A}+\mathbf{K}_{\mu}+\bm{\Sigma}_\mathrm{A}&\mathbf{K}_\mathrm{AB}&\mathbf{K}_\mathrm{AC}\\
		\mathbf{K}_\mathrm{AB}^\top&\mathbf{K}_\mathrm{B}+\mathbf{K}_{\mu}+\bm{\Sigma}_\mathrm{B}&\mathbf{K}_\mathrm{BC}\\
		\mathbf{K}_\mathrm{AC}^\top&\mathbf{K}_\mathrm{BC}^\top&\mathbf{K}_\mathrm{C}+\mathbf{K}_{\mu}+\bm{\Sigma}_\mathrm{C}\\
	\end{bmatrix}.
\end{equation}

In summary, the flexknot $m_\mathrm{fk}$ has $2n_\mathrm{fk}-1$ parameters, 
the offsets and time delays add $2(n_\mathrm{lc}-1)$ parameters, 
the white noise term adds one parameter and the microlensing kernels add two parameters since the \gls{se} kernel has two hyperparameters. The \gls{gp} kernel contributes its own parameters, the number of which depends on the kernel family (Appendix~\ref{appendix:kernel-choices}).

The priors on these parameters are summarised in Table~\ref{tab:priors}. In general, the prior ranges were chosen to depend only on the bounding data, i.e. the minimum and maximum extent of the light curves along the $t$- and $y$-axes, or data averages, as this sets the rough scale of the data without explicitly using all data points.
In practice, this achieves that the priors are both sufficiently wide and reasonably uninformative.
A drawback of wide uninformative priors is that it becomes computationally more expensive to converge to the dominant mode of the posterior, as it occupies a smaller fraction of the parameter space. We will see in Section~\ref{sec:protocol-nested-sampling-convergence} that incorporating more \emph{a priori} knowledge on the time delay parameter indeed accelerates the \gls{ns} run.

\begin{table}[!]
	\centering
	\caption{Priors set on the parameters of the flux variability model. All priors are uniform in the ranges shown. ``Sorted'' indicates that a sorted-uniform prior~\cite{buscicchio2019label, handley} was used. The top set of parameters corresponds to the time delays, the middle set to the mean function and the bottom set to the covariance matrix. The prior on the time delay $\Delta t$ was set differently for the \wfi and \bone $(1)$ and \he $(2)$ data sets, respectively. The range of $t$-values of the light curves is denoted by $\Delta t_\mathrm{range}=t_{n_\mathrm{data}}-t_1$, mean and std denote the mean and standard deviation of time delays reported in~\citet{hojjati2013robust}, $\bar{\mathbf{y}}_\mathrm{A}$ is the average of the magnitudes $\mathbf{y}_\mathrm{A}$ of light curve A and $f_\mathrm{Nyquist}$ is the Nyquist frequency computed from $t_1,\dots, t_{n_\mathrm{data}}$.}
	\label{tab:priors}
		\begin{tabular}{@{}ll@{}}
			\toprule
			\toprule
			Parameter & Prior range\\
			\midrule
			(1) $\Delta t_{\mathrm{AB}}$, $\Delta t_{\mathrm{AC}}$, $\Delta t_{\mathrm{AD}}$
			&$[-\Delta t_\mathrm{range},\Delta t_\mathrm{range}]$\\
			(2) $\Delta t_{\mathrm{AB}}$, $\Delta t_{\mathrm{AC}}$, $\Delta t_{\mathrm{AD}}$
			&$[\mathrm{mean}- 10\cdot \mathrm{std},\mathrm{mean}+10\cdot \mathrm{std}]$\\
			\midrule
			$\Delta m_{\mathrm{A}j}$ for $j\in \{\mathrm{B},\mathrm{C},\mathrm{D}\}$
			&$[\min_i(\mathbf{y}_j-\bar{\mathbf{y}}_\mathrm{A})_i,\max_i(\mathbf{y}_j-\bar{\mathbf{y}}_\mathrm{A})_i]$\\
			$n_\mathrm{fk}$&$\{0,1,\dots, 6\}$\\
			$c_\mathrm{fk}$
			&$[\min_i(\mathbf{y}_\mathrm{A}-\bar{\mathbf{y}}_\mathrm{A})_i,\max_i(\mathbf{y}_\mathrm{A}-\bar{\mathbf{y}}_\mathrm{A})_i]$\\
			$y_{1}^{(\mathrm{fk})},\dots,y_{n_\mathrm{fk}}^{(\mathrm{fk})}$
			&$[\min_i(\mathbf{y}_\mathrm{A}-\bar{\mathbf{y}}_\mathrm{A})_i,\max_i(\mathbf{y}_\mathrm{A}-\bar{\mathbf{y}}_\mathrm{A})_i]$\\
			$t_2^{(\mathrm{fk})},\dots,t_{n_\mathrm{fk}-1}^{(\mathrm{fk})}$&$\mathrm{Sorted}([t_1,t_{n_\mathrm{data}}])$\\
			\midrule
			$A$, $w_q$, $A_\mu$&$[0,\max_i\sqrt{\frac{\max_j (\mathbf{y}_i)_j - \min_j (\mathbf{y}_i)_j}{2}}]$\\
			$\ell$, $\ell_q$, $\ell_\mu$&$[0,\Delta t_\mathrm{range}]$\\
			$\alpha$&$[0,10]$\\
			$f_q$&$[0,f_\mathrm{Nyquist}]$\\
			$\sigma$&$[0,\max_{i,j}(\bm{\sigma}_i)_j]$\\
			\bottomrule
			\bottomrule
		\end{tabular}
\end{table}

\subsection{Marginalised Gaussian processes}\label{sec:marginalised-gps}

The flexibility of our \gls{gp} model is represented by the set of choices for the kernel and the mean function. A drawback of this is that we could potentially overfit the light curves. This motivates us to use Bayesian inferrence, wherein 
the Bayesian evidence selects the most probable model. Taking a step further, it is possible to marginalise over all such choices using marginalised \glspl{gp}~\citep{kroupa2024kernel, simpson}.
In comparison to traditional \glspl{gp}, they infer the joint posterior of the discrete choice over different mean functions, kernels or noise models as well as \gls{gp} kernel parameters. 
This is done by first inferring the posterior over \gls{gp} parameters using Bayes theorem, 
\begin{equation}\label{eqn:hyperparameter-posterior}
	p(\bm{\theta} \mid \mathcal{D}, \mathcal{M})=\frac{p(\mathcal{D} \mid \bm{\theta}, \mathcal{M})p(\bm{\theta}\mid \mathcal{M})}{p(\mathcal{D}\mid \mathcal{M})},
\end{equation}
where $\mathcal{D}$ is a data set, $\bm{\theta}$ are the parameters and $\mathcal{M}$ is the model choice,
and using the computed evidences ${p(\mathcal{D}\mid \mathcal{M})}$ to infer the posterior over the choices,
\begin{equation}\label{eqn:choice-posterior}
	p(\mathcal{M} \mid \mathcal{D})=\frac{p(\mathcal{D} \mid \mathcal{M})p(\mathcal{M})}{p(\mathcal{D})}.
\end{equation}
In all of our calculations, the prior $p(\mathcal{M})$ on the model choice 
is chosen to be uniform. 

We note that this approach presents a significant generalisation of the standard method of maximising the likelihood ${p(\mathcal{D} \mid \bm{\theta}, \mathcal{M})}$ over the parameters (commonly called the hyperparameters in the context of \glspl{gp}), known as type II maximum likelihood~\cite{rasmussen}.

It is possible to perform the calculation of both Equation~\ref{eqn:hyperparameter-posterior} and~\ref{eqn:choice-posterior} in a single \gls{ns} run by appropriately augmenting the \gls{ns} parameter space with model choice parameters~\cite{kroupa2024kernel}. 
We take the alternative approach of separately computing the posterior and evidence for each choice (Equation~\ref{eqn:hyperparameter-posterior}) since proper sampling of the posterior requires manual choice-dependent tuning of \gls{ns} convergence parameters (Section~\ref{sec:real-data-method}) due to intrinsic pathologies of the time delay likelihood~\cite{kroupa2026global}.

\section{Method}\label{sec:real-data-method}

In this section, we describe 
the method for sampling the posteriors. Finally, we explain how the inference of the time delays is stabilised with respect to the aforementioned pathologies in the likelihood.

\subsection{Sampling}

For each combination of flexknot mean function and kernel choice, a \gls{ns} run is performed for the quasars \wfi~\citep{vuissoz2008cosmograil} (${n_\mathrm{data}=218}$, ${n_\mathrm{lc}=3}$), \bone~\citep{fassnacht2002determination} (${n_\mathrm{data}=228}$, ${n_\mathrm{lc}=4}$) and \he~\citep{bonvin} (${n_\mathrm{data}=884}$, ${n_\mathrm{lc}=4}$). 
In practice, the \textsc{PolyChord}~\citep{handley2015polychord, handley} implementation of \gls{ns} is used.
Since the runtime of a \gls{ns} run is proportional to the likelihood evaluation time~\cite{petrosyan2023supernest, lemos2024improving}, the \gls{ns} runtime scales here as $\mathcal{O}((n_\mathrm{data}n_\mathrm{lc})^3)$~\cite{rasmussen}.
In practice, a single run typically took approximately $10^2$ CPU hours for the smallest data set, \wfi, to $10^3$ CPU days for the largest data set, \he.
Evidences and parameter posteriors are computed by post-processing samples with \textsc{anesthetic}~\cite{handley-anesthetic}. 
The flux variability model was implemented with the~\textsc{tinygp} library~\citep{tinygp}.

We sample the posterior with \gls{ns}~\cite{skilling}, primarily for three reasons. Firstly, \gls{ns} naturally incorporates uniform priors and priors with hard constraints, such as those from the flexknots, as \gls{ns} can operate without gradients of the likelihood.
Secondly, the \gls{gp} likelihood with a joint uniform prior on all hyperparameters may include singular covariance matrices which lead to divergent negative log likelihoods, i.e. zero likelihood. \gls{ns}, in particular \textsc{PolyChord}, adaptively ignores such regions without changing the evidence.
Thirdly, kernel, flexknot and time delay posteriors are observed to be typically multimodal and exhibit extended degeneracies~\cite{kroupa2026global}, which global methods such as \gls{ns} handle automatically.

\subsection{Protocol for Nested Sampling convergence}\label{sec:protocol-nested-sampling-convergence}

We observe that the time delay inference of all data sets suffers from the instability in~\citet{kroupa2026global}, i.e. the time delay posterior is peaked at values comparable to the width of the observation window. Notably, this occurs even for non-stationary mean functions, namely for flexknots with a large number of knots. This suggests that the structure of the likelihood described in~\citet{kroupa2026global} persists even if the assumption of stationarity is relaxed.

To mitigate this problem, we adopt the following protocol for the \wfi and \bone light curves. First, the individual \gls{ns} runs are performed with default \textsc{PolyChord} settings. At this stage most runs already produce time delays consistent with previously published values. Any runs for which the time delays are of order $\mathcal{O}(10^3\,\days)$ are rerun with the number of live points increased by a factor of $10$. A second iteration of this (increasing the number of live points by a total factor of $100$) is found to be sufficient to ensure that the time delays of all but one mean function and kernel combination are consistent within the error.
We hypothesise that the remaining unconverged run (four knots with an $\mathrm{SM}_{1}$ kernel for \wfi) did not converge due to mode evaporation. That is, if multiple modes are populated with live points during a~\gls{ns} run, it is possible for one or more modes to spontaneously lose their live points to another mode (``evaporate'') due to stochastic fluctuations in the live point population per mode. 
This can be mitigated by further increasing the number of live points, however this is costly in terms of the runtime~\cite{petrosyan2023supernest, lemos2024improving} of \gls{ns}.
Changing the random number generator seed, however, resulted in convergence with default \textsc{PolyChord} settings, showing that \gls{ns} is overall unstable with respect to this likelihood. 
For both \wfi and \bone, the prior on the time delay is uniform on the range $\pm \Delta t_\mathrm{range}$.

For \he, we observe that all \gls{ns} runs initially yield large time delays. This is unchanged under the above protocol,
within computational limitations. 
This is explained by noting that
the data set of \he is significantly larger.
Therefore, the posterior must be significantly more narrow around the true time delay when measured in terms of fractional parameter space volume. Due to the exceeding computational cost of increasing the number of live points further, we instead use a narrow uniform prior on the time delays. The prior is centred on the mean time delays inferred in \citet{hojjati2013robust} and widths set to $10$ times the reported standard deviation. 
We note that approximate evidence values for the full-width prior can be obtained by appropriate rescaling, assuming that the posteriors are approximately zero outside the narrow prior. The rescaling factor is the same for all models here and the posterior probabilities of the models are therefore unaffected by such a rescaling. Therefore, we do not apply this correction.
We find that inference with the narrow priors recovers well-constrained time delays 
with default \textsc{PolyChord} settings. 
Further, as the likelihood evaluation time is significantly larger, we fit a smaller number of models to the data, in particular limiting the maximum order of the Spectral Mixture kernels to \gls{sm}$_2$.

\section{Results}\label{sec:results}

\begin{figure*}
	\centering
	\includegraphics{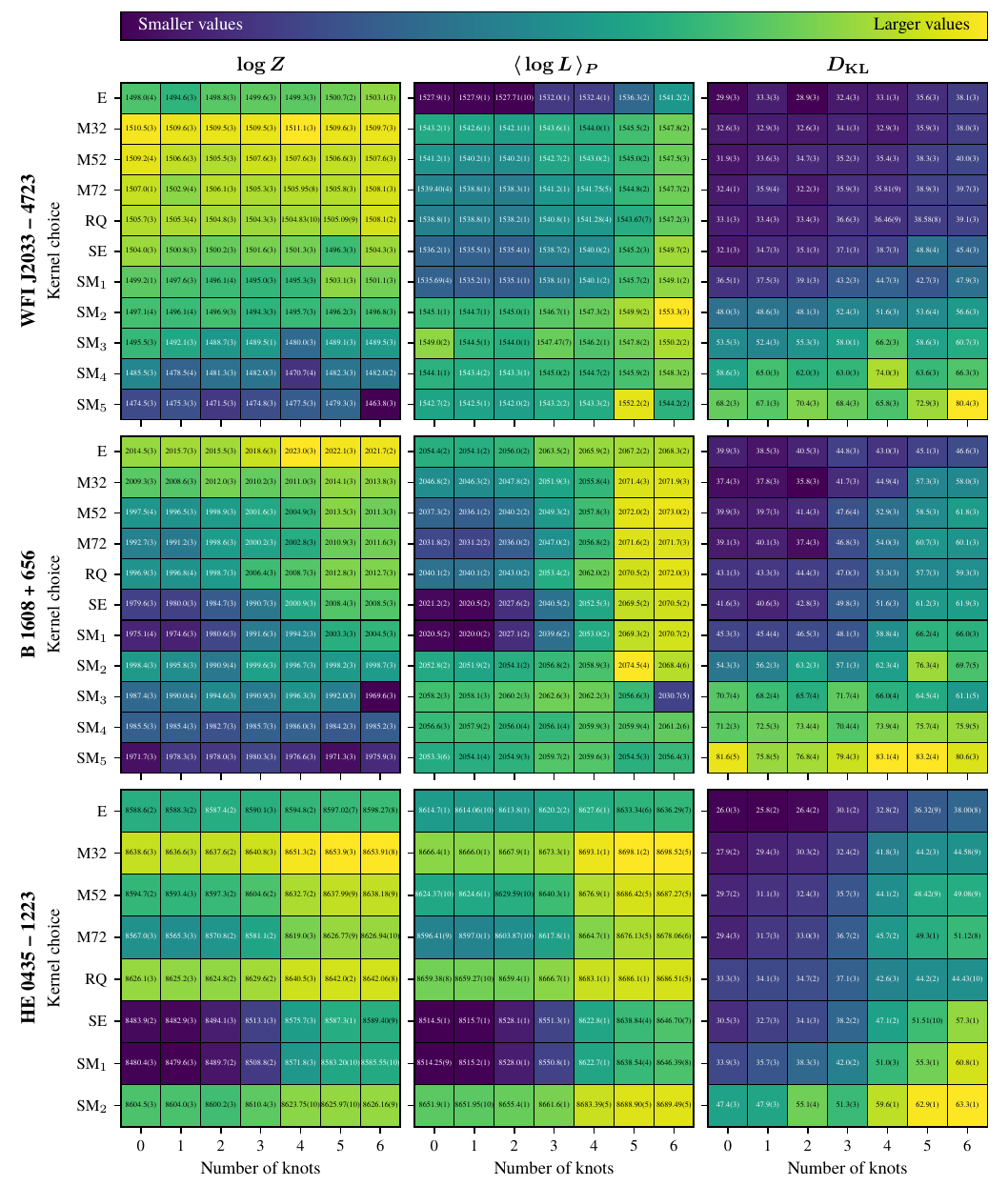}
	\caption{Evidence $\log Z$, posterior-averaged log-likelihood $\langle\log L\rangle_P$ and \gls{kl}-divergence $D_\mathrm{KL}$ for the three quasar light curve data sets. The three quantities are related by Occam's razor equation, $\log Z=\langle\log L\rangle_P- D_\mathrm{KL}$.
		Each coloured square in a subplot corresponds to a \gls{gp} fit, performed with a given number of knots $n_\mathrm{fk}$ in the flexknot mean function and a given \gls{gp} kernel. Zero and one knots correspond to a fixed and constant mean (with the constant a parameter), respectively. Note that the numbers in brackets denotes the standard deviation on the last digits.
		As shown by the \gls{kl}-divergence, model complexity increases roughly from the top left to the bottom right corner. The goodness of fit, $\langle\log L\rangle_P$, roughly shows the opposite trend, with more complex models producing better fits.
		As the balance of these two, the evidence $\log Z$ shows an overall preference to the \gls{e} and \gls{m32} kernels. Non-stationarity (more than one knot) is preferred for \bone and \he, whereas there is no clear indication of non-stationarity for \wfi.
	}
	\label{fig:real-data-evidences}
\end{figure*}

\begin{figure*}
	\centering
	\includegraphics{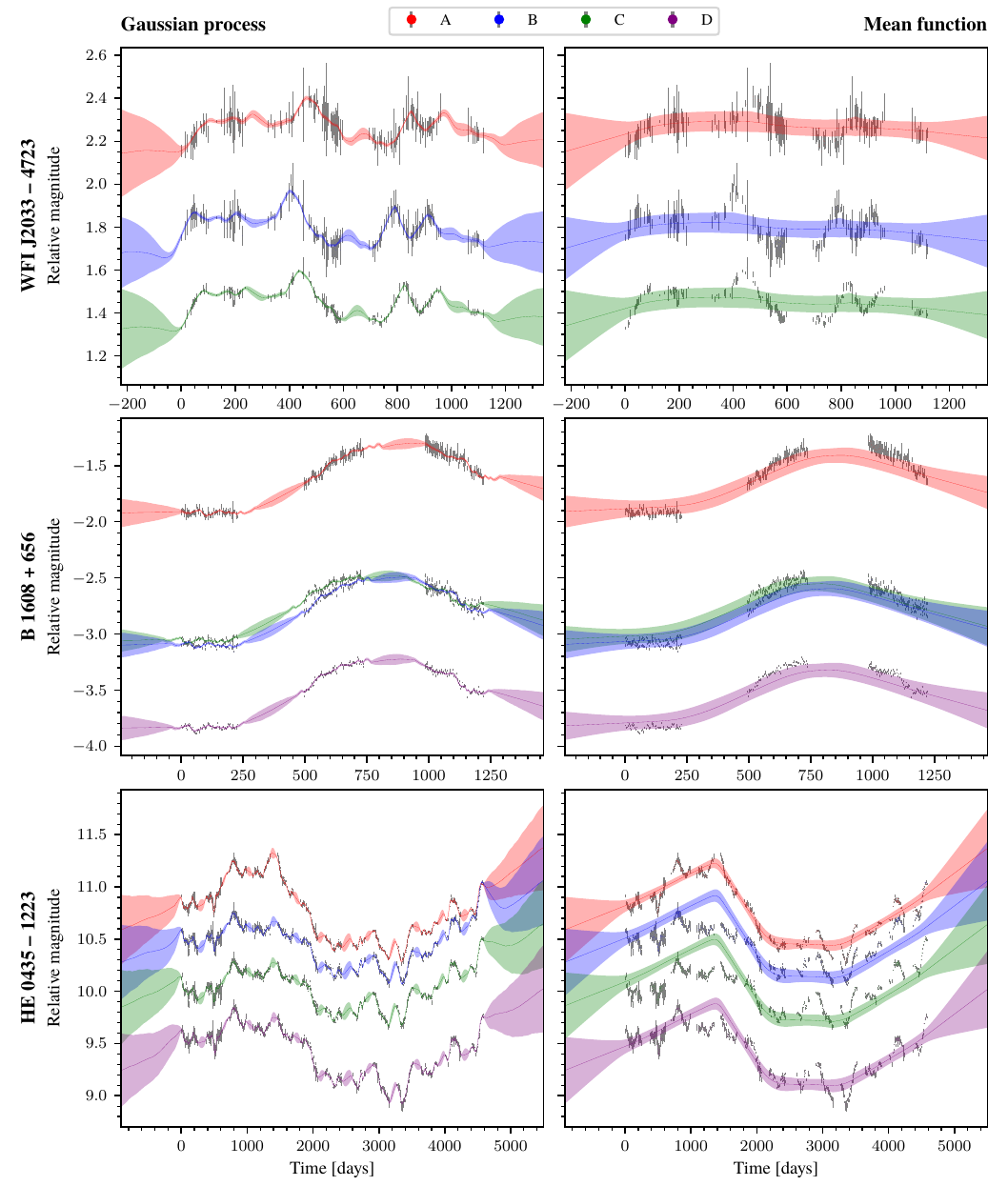}
	\caption{
		\gls{gp} (left column) and mean function (right column) posterior predictive distributions for the three quasar light curve data sets, each consisting of a number of light curves: A, B, C and (if existing) D. The \gls{gp} is the sum of the deterministic mean function and the stochastic contribution from the \gls{gp} kernel, which are fit jointly to the data. Plotting just the mean function therefore extracts the non-stationarity of the \gls{gp}. The fits are marginalised over all choices in Figure~\ref{fig:real-data-evidences}.
		Visually, the mean function of \wfi is consistent with a constant, whereas for \bone and \he, the mean function is clearly non-stationary.
		Note that the full \gls{gp} on the right reverts to the mean function outside the observation window, i.e. extrapolation is dictated by the mean function.
		The time axis is set to zero at the first observation.
	}
	\label{fig:gp-posterior-plots}
\end{figure*}

The following results were all obtained by regressing the light curves in units of relative magnitude instead of flux, as supported by previous work~\cite{uttley2001flux, uttley2005non}.
As an independent test, we also refitted our \gls{gp} model on the \wfi data set in units of flux, from which we obtained evidence values less than $-350$, which are strongly disfavoured compared to the fits in units of magnitude. Thus, \wfi prefers a \gls{gp} model which is log-normal in units of flux, and we proceed with fits in units of magnitude.

\subsection{Kernel and mean function inference}

\subsubsection{Kernel inference}

The posterior over mean function and kernel choices in terms of the model evidence $\log Z$ is shown in Figure~\ref{fig:real-data-evidences} and the corresponding marginalised fits are shown in Figure~\ref{fig:gp-posterior-plots}.
\wfi, \bone and \he exhibit a preference to the M32, E and M32 kernel, respectively. 
The \gls{e} kernel corresponds to a damped random walk and this preference is consistent with general observations on quasar light curves~\cite{zu}.
The preference to the \gls{m32} kernel indicates that the underlying stochastic process of the light curve is non-Markovian, i.e. there is a memory of previous values.

To test this property more precisely, we 
restricted the \he data to some subset $t_1,\dots,t_{n_\mathrm{subset}}$ and 
computed the Bayes factor between the \gls{m32} and \gls{e} kernel for increasing $n_\mathrm{subset}$. The flux variability model was chosen to be a constant mean function model. This showed that the Bayes factor increases with observation duration $n_\mathrm{subset}$ and suggests that features in the light curves at longer time scales contribute significantly to the kernel posterior and that the Markov property of the \gls{e} kernel fails to hold on longer time scales. 

Finally, we note that 
the Mat\'ern kernels
can be further generalised to yield the \gls{carma} kernels~\cite{kelly2014flexible}. However, as our results show, the complexity penalty 
is already sufficient to substantially decrease the Bayesian evidence for $\nu=\frac52$. \gls{sm} kernels are generally disfavoured as well, so we do not consider this extension here. 
The power spectral density of a \gls{carma} kernel is a weighted sum of Lorentzians instead of Gaussians. It remains to be seen in future work whether the heavier tails of such a power spectral density yield substantially different results.

\subsubsection{Mean function inference}\label{sec:mean-function-inference}

Considering the inferred mean function, 
Figure~\ref{fig:real-data-evidences} shows that the evidence continues to increase with the number knots for \bone and \he and therefore clearly show a preference to a larger number of knots. 
The mean function posterior in Figure~\ref{fig:gp-posterior-plots} agrees with this, showing mean functions which follow the average fluctuation of the data.

We ascribe 
the mean function preferences to the overall structure of the light curves on longer time scales. In particular, the light curves of \bone and \he visibly contain variations of large amplitude on longer time scales, superimposed on smaller variations on shorter time scales. The mean function effectively separates out the long time-scale variations, leaving the fluctuations around the mean to be fit by the \gls{gp} kernel. Moreover, this implies that the short time-scale variations are well-described by the E and M32 kernel for \bone and \he, respectively.
As it could be possible for kernels with multiple length scales (\gls{rq} and \gls{sm} kernels) to fit the longer scale variations with one length scale and shorter scale variations with another, the fact that single length scale kernels (\gls{e} and \gls{m32} kernels) are preferred indicates strongly that the variation modelled by the flexknot is indeed a deterministic drift in the light curve.

In contrast, visual inspection of Figure~\ref{fig:gp-posterior-plots} shows that the mean function of \wfi is approximately constant and agrees with the empirical mean of each light curve. 
We would therefore expect that the model evidence decreases as the number of knots is increased.
However, the number of knots remains unconstrained in Figure~\ref{fig:real-data-evidences}, i.e. the model posterior is approximately flat as the number of knots is increased, for the most probabable kernel \gls{m32}.
Furthermore, $\langle\log L\rangle_P$ and $D_\mathrm{KL}$ both increase at the same rate with increasing number of knots.
Overall, this implies that higher order flexknots significantly contribute to the mean function posterior, but exhibit only little variation around a constant mean.

To explain the behaviour of the evidence, we first recall that we use a sorted-uniform prior on the $t$-values of the flexknot. Therefore, as we add more knots, each knot can only access an interval of width $\frac{\Delta t_\mathrm{range}}{n_\mathrm{fk}}$ on the $t$-axis, on average, times an fixed interval along the $y$-axis. That is, the accessible parameter space per knot is a rectangle, on average. Clearly, the width of this rectangle goes to zero as the number of knots becomes large. In this sense, the flexknot is self-constraining, namely adding more parameters does not exponentially increase the accessible parameter space.
To offer a more precise explanation, the total parameter space accessible to the $t$-positions of the knots is $\frac{\Delta t_\mathrm{range}^{n_\mathrm{fk}}}{n_\mathrm{fk}!}$~\cite{handley2015polychord, buscicchio2019label}, which goes to zero as the number of knots $n_\mathrm{fk}$ becomes large. For such models in which the prior becomes more narrow with an increasing number of parameters, it is possible that the evidence flattens out beyond a certain threshold value of the model complexity~\cite{rasmussen2000occam}. 
The genericity of the above argument is supported by the same phenomenon observed on flexknots in a different context~\cite{ormondroyd2025nonparametric, ormondroyd2025comparison}.
We leave the question of whether this problem is resolved by a change of prior (for example a sparsity promoting prior) to future work.

\subsubsection{Goodness of fit and model complexity}

Figure~\ref{fig:real-data-evidences} further shows the posterior-averaged log-likelihood $\langle\log L\rangle_{P}$ as a measure of the goodness of fit and the \gls{kl} divergence $D_\mathrm{KL}$ as a measure of model complexity. These are related to the evidence by $\log Z = \langle\log L\rangle_{P} - D_\mathrm{KL}$~\cite{hergt}.
The model complexity
increases with the number of knots, as expected.
We also observe an increase with the order of the Mat\'ern kernel and the Spectral Mixture kernel. The former corresponds to increasing smoothness of the corresponding \gls{gp}. 
That is, the model complexity roughly increases from top to bottom on our kernel axis.
This ordering of kernels in terms of complexity is consistent with previous results for \glspl{gp}~\cite{kroupa2024kernel}. 

The \gls{kl} divergence of \he has little effect on the model evidence. That is, the fit is primarily driven by the goodness of fit. In contrast, the \gls{kl} divergence enacts a notable regularisation for \wfi and \bone. This behaviour is expected as the data set of \he is significantly larger. More precisely, we expect the posterior-averaged log-likelihood and \gls{kl} divergence to scale approximately linearly and logarithmically with the number of data points, respectively, so that the latter becomes subdominant~\citep{trotta}.

\subsubsection{Preference to the reference light curve}\label{sec:preference-to-the-reference-light-curve}

We note that the mean function of \he appears to fit light curve A visually better than the others. Specifically, the magnitude of light curve A increases more strongly around $t=1400\,\days$ 
compared to the others, and the mean function follows this increase. This is certainly unexpected, as the penalty from the mis-fit to the three other light curves should outweigh the goodness of fit to light curve A. 
The underlying problem here is that the light curve model treats light curve A as a reference light curve (Section~\ref{sec:flux-model-mean-function}). That is, other light curves are modelled as shifted and distorted versions of light curve A. Moreover, the priors (Table~\ref{tab:priors}) are set by the properties of light curve A, for instance the range of the flexknot knot $t$-values. We confirmed that the mean function fits the reference light curve better by refitting with a different choice of reference light curve on synthetic data.

Clearly, the choice of light curve A as the reference light curve is artificial and ideally the light curve model should be permutation-invariant to the relabelling of light curves and show no affinity to a particular light curve. An \textit{a posteriori} solution is therefore to perform multiple fits for each choice of reference light curve and subsequently average over this choice following Bayes theorem. More precisely, it is possible to treat the choice of the reference as a further parameter on which we could set a uniform prior, calculate the evidence for each choice, which in turn is obtained from the evidences of all flexknot and kernel choices, and finally marginalise over these choices. We do not proceed with this solution due to the increase in computational cost and the fact that the mean functions at least visually recover the correct trend for all light curves.
Instead, the formulation of a permutation-invariant flux variability model is left to future work.

\subsection{Stationarity}

From the results described in Section~\ref{sec:mean-function-inference}, it follows that the stochastic process of the \wfi light curves is consistent with stationarity, whereas the \bone and \he light curves are non-stationary. 
Note however that the mean functions of \bone and \he do not exhibit an overall drift to larger or smaller magnitudes but instead first increase and subsequently decrease in an oscillatory manner. It is therefore still possible that the light curve is stationary but exhibits stochastic variations on multiple length scales when measured for a much longer observation window. 
In this case, we would expect that the evidence values shift to favour the \gls{rq} or \gls{sm} kernels, which contain multiple length scales, compared to the \gls{e} and \gls{m32} kernel, only contain a single length scale.
For the given data, we therefore conclude that the light curves of \bone and \he are indeed non-stationary within the adopted flux variability model.

\subsection{Time delays}

\begin{figure*}[!]
	\centering
	\includegraphics{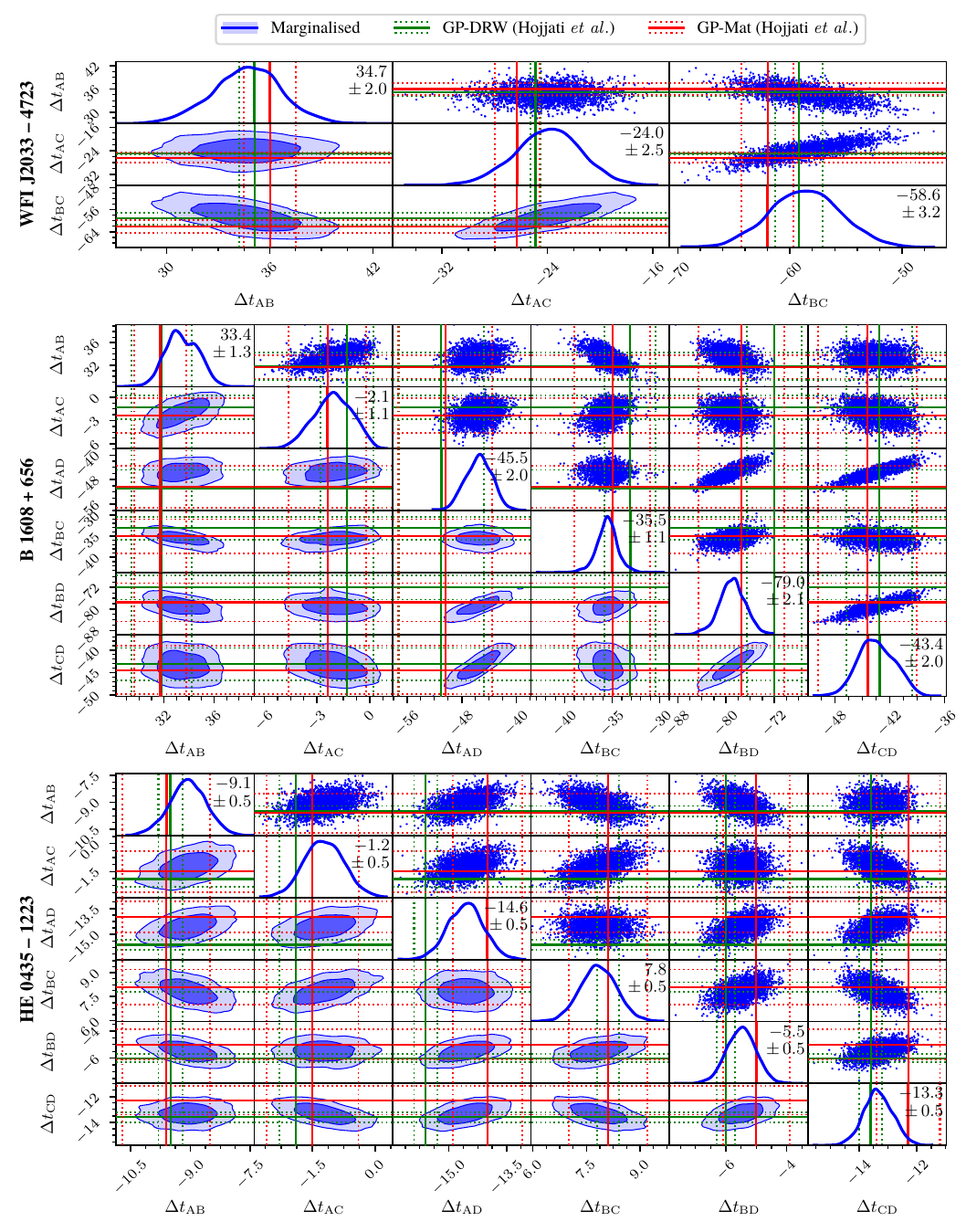}
	\caption{
		Corner plots of the time delay posteriors for the three data sets. The posteriors are marginalised over all model choices in Figure~\ref{fig:gp-posterior-plots}. For comparison, the inferred values (mean and standard deviation) from \citet{hojjati2013robust} are overplotted. All posteriors are unimodal and well-constrained, with the exception of $\Delta t_\mathrm{AB}$ of \bone, which displays a weakly protruding secondary peak. The numerical means and standard deviations are also shown in Table~\ref{tab:timedelays}.
	}
	\label{fig:time-delay-corner-plots}
\end{figure*}

\begin{table*}[!]
	\centering
	\caption{Comparison of the time delay posterior values for the quasars \wfi, \bone and \he (Figure~\ref{fig:time-delay-corner-plots}) with those from \citet{hojjati2013robust}. The values were calculated by marginalising over the model choices in Figure~\ref{fig:gp-posterior-plots}. The results in \citet{hojjati2013robust} were obtained for the exponential (DRW) and Mat\'ern-3/2 kernels. All time delays are in units of days.}
	\label{tab:timedelays}
		\begin{tabular}{@{}lcccccc@{}}
			\toprule
			\toprule
			System and Method & $\Delta t_\mathrm{AB}$ & $\Delta t_\mathrm{AC}$ & $\Delta t_\mathrm{AD}$ & $\Delta t_\mathrm{BC}$ & $\Delta t_\mathrm{BD}$ & $\Delta t_\mathrm{CD}$ \\
			\midrule
			WFI J2033--4723 Marginalised & \timedelayval{34.7}{1.9} & \timedelayval{-24.0}{2.5} & \missingdata & \timedelayval{-58.7}{3.1} & \missingdata & \missingdata \\
			WFI J2033--4723 GP-DRW \citep{hojjati2013robust} & \timedelayval{35.1}{0.9} & \timedelayval{-24.9}{0.4} & \missingdata & \timedelayval{-59.2}{2.1} & \missingdata & \missingdata \\
			WFI J2033--4723 GP-Mat \citep{hojjati2013robust} & \timedelayval{36.0}{1.5} & \timedelayval{-26.3}{1.7} & \missingdata & \timedelayval{-62.0}{2.3} & \missingdata & \missingdata \\
			\midrule
			B1608+656 Marginalised & \timedelayval{33.4}{1.3} & \timedelayval{-2.1}{1.1} & \timedelayval{-45.4}{2.0} & \timedelayval{-35.5}{1.1} & \timedelayval{-78.8}{2.1} & \timedelayval{-43.3}{2.0} \\
			B1608+656 GP-DRW \citep{hojjati2013robust} & \timedelayval{31.8}{2.4} & \timedelayval{-1.3}{1.5} & \timedelayval{-51.0}{6.2} & \timedelayval{-33.1}{2.7} & \timedelayval{-72.0}{4.5} & \timedelayval{-43.1}{3.6} \\
			B1608+656 GP-Mat \citep{hojjati2013robust} & \timedelayval{31.7}{2.1} & \timedelayval{-2.4}{2.2} & \timedelayval{-50.4}{6.9} & \timedelayval{-35.0}{4.0} & \timedelayval{-77.5}{7.1} & \timedelayval{-44.4}{5.4} \\
			\midrule
			HE 0435--1223 Marginalised & \timedelayval{-9.1}{0.5} & \timedelayval{-1.2}{0.5} & \timedelayval{-14.5}{0.5} & \timedelayval{7.8}{0.5} & \timedelayval{-5.5}{0.5} & \timedelayval{-13.3}{0.5} \\
			HE 0435--1223 GP-DRW \citep{hojjati2013robust} & \timedelayval{-9.5}{0.3} & \timedelayval{-1.9}{0.4} & \timedelayval{-15.6}{0.3} & \timedelayval{8.1}{0.3} & \timedelayval{-6.0}{0.3} & \timedelayval{-13.6}{0.4} \\
			HE 0435--1223 GP-Mat \citep{hojjati2013robust} & \timedelayval{-9.6}{1.1} & \timedelayval{-1.5}{1.1} & \timedelayval{-14.0}{0.9} & \timedelayval{8.1}{1.1} & \timedelayval{-5.0}{1.1} & \timedelayval{-12.3}{1.1} \\
			\bottomrule
			\bottomrule
		\end{tabular}
\end{table*}

Our fits directly yield joint samples of the time delays $\Delta t_\mathrm{AB}$, $\Delta t_\mathrm{AC}$ and $\Delta t_\mathrm{AD}$ measured relative to light curve~A. This allows us to calculate samples of the other time delays, $\Delta t_\mathrm{BC}$, $\Delta t_\mathrm{BD}$ and $\Delta t_\mathrm{CD}$, by taking appropriate differences.
The resulting joint posteriors with mean and standard deviation 
are shown in Figure~\ref{fig:time-delay-corner-plots} and Table~\ref{tab:timedelays}. 

We compare with the results in \citet{hojjati2013robust}, in which the inference was performed with a constant mean function and an E (damped random walk) or M32 kernel. 
To ensure a direct comparison under matched modelling assumptions, we use the subset of our model grid which exactly matches their models (number of knots $n_\mathrm{fk}=1$, and either an E or M32 kernel). No additional fitting is required for this comparison as these are \gls{ns} runs already included in our analysis, and are distinct from the marginalised results reported in Table~\ref{tab:timedelays}.
We find that while the mean time delays change, they agree within the error, i.e. the time delay
does not strongly depend on the choice of the kernel.
Beyond this, our results show that time delay inference is broadly robust under the choice of mean function. Overall, we do not observe large tensions between individual model choices.

For \wfi, the error bars of our fits increase. 
We rule out that this is due to marginalisation over different models. Indeed, by inspecting our results purely for the E and M32 kernel with a single knot in the mean function, we validate that these time delays are even larger than the marginalised time delays. Marginalisation therefore reduces the error in this case.
For the E kernel, the length scale posterior shows a substantial degeneracy in parameter space. That is, the posterior is flat above a certain length scale. However, we rule out that this causes the larger time delay error bars since the length scale posterior of the M32 kernel, which has a higher evidence, is properly constrained.
It is therefore possible that the difference may arise from the lack of convergence of the sampling method used in \citep{hojjati2013robust}, which we cannot reproduce here and therefore do not further investigate.
Finally, we note that \wfi generally has a broad time delay posterior, as discussed in Section~\ref{sec:real-data-method}, which we ascribe to the small data set size relative to the other two data sets.

For \bone and \he, the size of the error bars is smaller than and lies in between those of the E and M32 kernel, respectively. The error bars obtained from our time delay posteriors for only the E and M32 kernels is of comparable size to the marginalised ones. As above, it is possible that this discrepancy arises from the convergence of the sampling methods used. 

It is noteworthy that we use an updated version of the \he data set, which is approximately twice as large than the earlier one used in~\citet{hojjati2013robust}. Despite this increase in data set size, the time delays only change within the error. It is therefore possible that future measurements may not significantly influence the time delays further but instead predominantly inform the modelling of the light curve itself. However, a rigorous investigation of the influence of more data is beyond the scope of this work.

\subsection{Scatter around the mean}

\begin{figure}[!]
	\centering
	\includegraphics{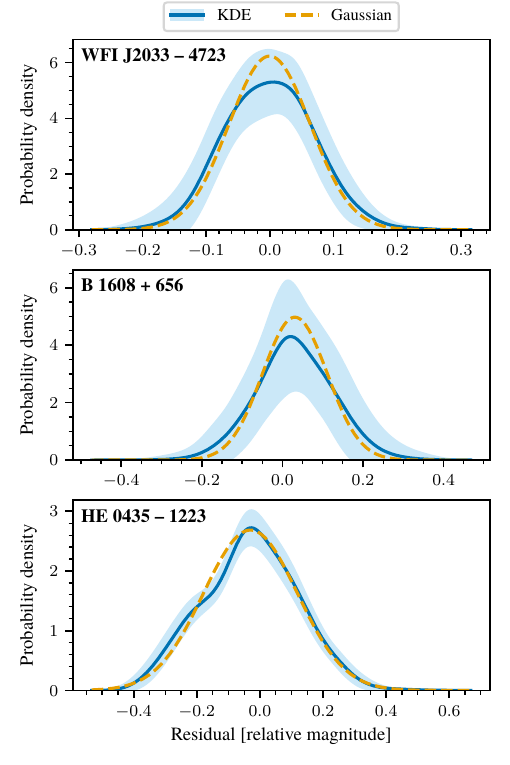}
	\caption{Scatter of the light curve data points around the mean function of the \gls{gp} fit in units of magnitude.
		The distribution of the residuals is expected to follow a normal distribution. 
		For comparison, a Gaussian distribution with the same mean and standard deviation is shown as well. For \wfi and \bone, the agreement in the tails is better than in the bulk. For \he, the observed deviation from the exact Gaussian on the left half is attributed to a mis-fit resulting from lack of permutation-invariance of the \gls{gp} model.}
	\label{fig:scatter-around-mean}
\end{figure}

For a stationary \gls{gp} $y(t)$, we expect the histogram of the observed $y$-values to converge to a Gaussian distribution. This is an immediate consequence of the fact that stationarity, by definition, implies that the marginal distribution $p(y)$ is the same for all times $t$. However, an important assumption here is that the observations of the \gls{gp} are representative of the true underlying distribution. This is only guaranteed if the observation time becomes sufficiently long so that ergodicity applies~\citep{pavliotis2014stochastic, grimmett2020probability}. 
The constraint of ergodicity is only satisfiable by measuring a light curve for a long time and can therefore strictly not be resolved by any method of inference. 

Nevertheless, we can subtract the inferred non-stationary mean function from the data to render our \gls{gp} model stationary and inspect the resulting histogram of residuals. 
Historically~\cite{uttley2001flux, uttley2005non}, the observation that the flux distribution is log-normal, i.e. that magnitudes are Gaussian, provided one piece of evidence that the underlying processes in the accretion disc follow a multiplicative stochastic process instead of an additive one.
Confirmation that the residuals with respect to our model possess a Gaussian distribution in units of magnitude therefore poses an important consistency test of our modelling assumptions.

Since we do not have a single mean function but instead a posterior distribution over mean functions, we can sample a mean function, calculate the residuals, plot a \gls{kde} and repeat this for every posterior sample. Thus, we obtain a distribution over \glspl{kde}, which we can plot by calculating the mean and error bar over \glspl{kde}.

Figure~\ref{fig:scatter-around-mean} shows the mean and error bar of the \gls{kde} for \wfi, \bone and \he. 
For comparison, we also show a normal distribution with mean and standard deviation set by the mean and standard deviation of the residual samples. 
Visually, all three residual distributions are consistent with normal distributions, within the error. Deviations around the peak of the normal distribution from the mean \gls{kde} are larger, and the tails of the normal distribution are lighter. 
These deviations are within one standard deviation.

For \he, the error bars around the \gls{kde} are much narrower than for the other two data sets, which we attribute to the larger size of the data set. Similarly, the error bars on the time delays are smallest for \he. While the right tail agrees well with the normal distribution, the left tail does not decay smoothly but displays an inflection point which causes larger deviation from the tail of the normal distribution. This deviation is consistent with the mis-fit of the mean function discussed in Section~\ref{sec:preference-to-the-reference-light-curve}, visible in Figure~\ref{fig:gp-posterior-plots}. That is, we expect the mis-fit of the mean function at around $t=1400\,\days$ for light curves B, C and D to lead to a distortion in the distribution of the residuals.

Evidently, the uncertainty around the kernel density estimate is much larger for \bone compared to \wfi. Considering Figure~\ref{fig:gp-posterior-plots}, this is consistent with the fact that the uncertainty on the mean function is comparable to the fluctuations of the data around the mean. 
In contrast, the uncertainty of the mean function is much smaller for \wfi and \he, compared to the fluctuation of the data around the mean. The \gls{kde} of residuals for these two data sets therefore fluctuates less.

\subsection{Discussion of Nested Sampling instabilities}

We conclude the presentation of results by offering further convergence diagnostics for the nested sampling instabilities in the context of time delay inference, in addition to the convergence protocol in Section~\ref{sec:protocol-nested-sampling-convergence}.
We first stress that the evidence values of the unconverged and converged runs can differ by significantly more than five sigma, so ensuring convergence is crucial for an accurate model posterior. We also observe that unconverged runs visually appear as discontinuities in the $\log Z$, $\langle\log L\rangle_P$ and $D_\mathrm{KL}$ plots and proper convergence causes these values to change more smoothly when plotted as a function of model complexity, such as in Figure~\ref{fig:real-data-evidences}. This serves as a further heuristic for assessing convergence, beyond monitoring the time delay.
Finally, we have compared the marginalised time delays for both converged (obtained by our protocol) and unconverged (obtained with default \textsc{PolyChord} settings) runs for the \wfi data set and the resulting shift in the time delays lies well within one sigma. This robustness is due to the marginalisation over models, which protects the inference from spurious unconverged values (which have low evidence). We conjecture that the mechanism behind this is that, if the dominant peak in the posterior is not found by nested sampling, the evidence automatically decreases, so that the posterior probability of that model becomes smaller and therefore contributes less to the final marginalised inference. This mechanism is only helpful if there are few unconverged runs among many properly converged runs, unlike the inference for \he, for which no runs had initially converged (though adoption of the convergence protocol lead to proper convergence thereafter).

Finally, we stress that the \gls{ns} convergence problems fundamentally arise from the use of an insufficient number of live points, which itself is limited by the computational cost. Advances in sampling on GPUs~\cite{yallup2025nested, yallup2026nested} were shown to allow for a substantial increase in the number of live points at negligible additional cost, therefore offering a promising venue for future work.

\section{Conclusions}\label{sec:conclusions}

In this work, we have introduced a robust Bayesian framework for analysing gravitationally lensed quasar light curves by marginalising over a flexible set of \gls{gp} models. To disentangle long-term, non-stationary trends from intrinsic stochastic variability, we modelled the underlying light curve with a deterministic mean function, parametrised by a piecewise linear spline known as a flexknot, and a stationary stochastic component described by kernels from the Mat\'ern and Spectral Mixture families. This hierarchical approach allows for the inference of cosmological time delays that are robust to assumptions about the underlying variability model.

Applying this framework to the lensed quasars \wfi, \bone, and \he, our analysis reveals distinct variability properties for each system. We find strong evidence that the light curves of both \bone and \he are non-stationary, requiring a flexible mean function to model their long-term deterministic drifts. However, their underlying stochastic processes differ. \bone strongly prefers a Markovian Exponential kernel, while \he is best described by a non-Markovian Mat\'ern-$\frac32$ kernel. In contrast, the \wfi light curves are consistent with stationarity, yet their stochastic component also shows a strong preference for the non-Markovian Mat\'ern-$\frac32$ kernel. Our analysis also uncovered a model artifact, wherein the fit shows a preferential dependence on the arbitrarily chosen reference light curve.

The time delays inferred from our marginalised posteriors are consistent with previous findings, demonstrating the robustness of these measurements, but with the advantage of systematically accounting for uncertainty in the choice of the \gls{gp} kernel and mean function. Furthermore, by subtracting the inferred non-stationary drift, we show that the residuals of the light curve data are consistent with a Gaussian distribution, thereby validating a core assumption of the \gls{gp} model.

Methodologically, we highlight the critical importance of rigorous convergence testing for the nested sampling algorithm to ensure accurate Bayesian evidence calculation. We note, however, that the process of marginalising over a wide range of models provides a degree of robustness to the final time delay inference, mitigating the impact of unconverged results from any single model.

Looking forward, a natural extension of this work is to formulate a permutation-invariant \gls{gp} model to eliminate the artifact associated with the reference light curve. Another promising avenue is the extension to multi-wavelength data, where observations that trace bulk flows in the accretion disc could provide stronger physical constraints on the mean function, improving both the model's fidelity and its extrapolative power. We envision that this work lays a solid foundation for a more comprehensive and fully Bayesian approach to time delay inference, which, when paired with advances in \gls{gp} computation such as sparse \glspl{gp}, will pave the way for highly robust analyses of quasar light curves.

\section*{Acknowledgements}
N. K. gratefully acknowledges Andrew Fabian for discussions on stationarity in quasar light curves, Eric V. Linder and Alex G. Kim for assistance in reproducing \gls{gp} inference results, Christopher D. Fassnacht for providing the \wfi data and G\'abor Cs\'anyi and Christopher C. Lovell for helpful discussions.
N. K. was supported by the Harding Distinguished Postgraduate Scholarship. 
This work was performed using the Cambridge Service for Data Driven Discovery (CSD3), part of which is operated by the University of Cambridge Research Computing on behalf of the STFC DiRAC HPC Facility (www.dirac.ac.uk). The DiRAC component of CSD3 was funded by BEIS capital funding via STFC capital grants ST/P002307/1 and ST/R002452/1 and STFC operations grant ST/R00689X/1. DiRAC is part of the National e-Infrastructure.

\appendix

\section{Kernel choices}\label{appendix:kernel-choices}

The Mat\'ern-$\nu$ kernel~\citep{rasmussen}, indexed by $\nu$, is defined by
\begin{equation}\label{eqn:matern-kernel}
	k_\nu(\tau)=A^2\frac{2^{1-\nu}}{\Gamma(\nu)}\left(\frac{\sqrt{2\nu}}{\ell}\tau\right)^\nu K_\nu\left(\frac{\sqrt{2\nu}}{\ell}\tau\right),
\end{equation}
which has two hyperparameters, the length scale $\ell$ and amplitude $A$. Here, the kernel is only a function of the time difference $\tau=|t-t'|$ as the kernel is stationary. $K_\nu$ is a modified Bessel function of the second kind.
For half-integer $\nu=k+\frac12$, with integer $k$, the \gls{gp} is $k$-times differentiable in the mean-square sense and thus becomes more smooth as $\nu$ increases. The Mat\'ern-1/2 kernel is also known as the \gls{e} kernel and the limit $\nu\rightarrow\infty$ gives the \gls{se} kernel.

Equivalently, 
the \gls{gp} induced by the Mat\'ern kernel may be 
understood
as the solution to the following stochastic differential equation~\citep{hartikainen, sarkka2019applied}:
\begin{equation}\label{eqn:matern-sde}
	\left(\frac{\mathrm{d}}{\mathrm{d}t}+\frac{\sqrt{2\nu}}{\ell}\right)^{k+1}f(t)=\epsilon(t).
\end{equation}
Here, $\epsilon$ is a white noise process with mean zero and autocorrelation $\langle\epsilon (t)\epsilon(t')\rangle=q\delta(t-t')$ driving the stochastic process $f$. 
The noise strength $q$ is related to $\ell$ and $A$ by
\begin{equation}
	q=2 A^2\pi^{1/2}\frac{\Gamma(k+1)}{\Gamma(k+1/2)}\left(\frac{\sqrt{2\nu}}{\ell}\right)^{2k+1}.
\end{equation}
The Mat\'ern kernel is then the autocorrelation of the solution $f$ in steady-state, i.e. in the $t\rightarrow \infty$ limit.
This is consistent with the fact that we are not guaranteed to infer the true underlying kernel from given data if the observation time interval is short, regardless how fine the data is sampled in time, but only if it is sufficiently large compared to the length scale.

From Equation~\ref{eqn:matern-sde}, we see that increasing the index $\nu=k+\frac12$ introduces higher-order derivatives of $f$. Intuitively, this increases the smoothness of $f$ if $f$ is roughly thought of as the result of repeatedly integrating the white noise $\epsilon$. Moreover, as the length scale $\ell$ is increased, variations in $f$ occur over longer time scales as the weight of lower-order derivative terms decreases.
Increasing the amplitude~$A$, the strength of the white noise process driving $f$ increases, and therefore the variance of the distribution of $f$ increases as well, in agreement with Equation~\ref{eqn:matern-kernel} since $k_\nu(0)$ is equal to the variance.

As a final comment, 
the inclusion of second- or higher-order derivatives in Equation~\ref{eqn:matern-sde} introduces memory into the stochastic process.
This is readily seen from the counterpart in discrete time wherein the derivatives are replaced by time-shifts. In fact, for $\nu=\frac12$, we obtain the damped random walk, which is the unique stationary Markov Gaussian process~\citep{doob}. Larger values of $\nu$ are therefore non-Markov.

Another common choice for light curve modelling is the \gls{rq} kernel~\citep{wilkins2019low, griffiths2021modeling, covino2022detecting}, which is derived as a length scale mixture of \gls{se} kernels, 
\begin{equation}
	k_\mathrm{RQ}(\tau)=A^2\left(1+\frac{\tau^2}{2\alpha\ell^2}\right)^{-\alpha},
\end{equation}
which has three hyperparameters: the amplitude $A$, length scale $\ell$ and $\alpha$, which controls the heaviness of the kernel tail. As $\alpha\rightarrow\infty$, the \gls{se} kernel is recovered.

Since the Mat\'ern kernel only has a single length scale but quasar light curves were shown to exhibit multiple length scales
or periodicity, we also consider the \gls{sm}$_Q$ kernel, indexed by a positive integer~$Q$~\citep{wilson}. The motivation here is that any power spectral density can be approximated arbitrarily well by a sufficiently large mixture of Gaussians. The Fourier transform of such a power spectral density yields the kernel
\begin{equation}
	k_Q(\tau)=\sum_{q=1}^Q w_q\, \mathrm{e}^{-2\pi^2\tau^2/\ell_q^2}\cos(2\pi f_q\tau),
\end{equation}
where the weights $w_q$, length scales $\ell_q$ and frequencies~$f_q$ constitute $3Q$ hyperparameters in total. Further motivation for this kernel lies in the fact that the multiple length scales and periodicity may help to extrapolate better, thus addressing the extrapolation problem~\cite{kroupa2026global}.

In practice, we will use the \gls{e}, \gls{m32}, \gls{m52}, \gls{m72}, \gls{rq} and \gls{se} kernels and \gls{sm}$_Q$ kernels up to $Q=5$.

\bibliography{quasar_time_delays}

\end{document}